\documentclass[11pt]{article}
\usepackage{graphicx}
\usepackage{amsmath}
\usepackage{amssymb}
\usepackage{caption2}
\begin{document}
\bibliographystyle{prsty}
\begin{center}
{\large {\bf \sc{  Analysis of  the $\Lambda_c(2625)$ and $\Xi_c(2815)$   with QCD sum rules }}} \\[2mm]
Zhi-Gang Wang \footnote{E-mail:zgwang@aliyun.com.  }     \\
 Department of Physics, North China Electric Power University,
Baoding 071003, P. R. China
\end{center}

\begin{abstract}
In this article, we study   the
charmed baryon states $\Lambda_c(2625)$ and $\Xi_c(2815)$ with the spin-parity ${3\over 2}^-$
 by subtracting the contributions
from the corresponding    charmed  baryon states with the spin-parity ${3\over 2}^+$ using the QCD sum rules, and suggest a formula
$ \mu=\sqrt{M_{\Lambda_c/\Xi_c}^2-{\mathbb{M}}_c^2}$ with the effective mass  ${\mathbb{M}}_c=1.8\,\rm{GeV}$  to determine the energy scales of the QCD spectral densities, and make reasonable predictions for the  masses and pole residues.  The numerical
results  indicate that the  $\Lambda_c(2625)$ and $\Xi_c(2815)$ have at least two remarkable under-structures.
\end{abstract}



\section{Introduction}
In the past years,   several new charmed baryon states have been observed,  and    the
  spectroscopy of the charmed  baryon states have re-attracted  much attentions.
The ${1\over 2}^+$ and ${1\over 2}^-$ antitriplet charmed
baryon states ($\Lambda_c^+, \Xi_c^+,\Xi_c^0)$ and
($\Lambda_c^+(2595), \Xi_c^+(2790),\Xi_c^0(2790))$,  and the
${1\over 2}^+$ and ${3\over 2}^+$ sextet charmed baryon states
($\Omega_c,\Sigma_c,\Xi'_c$) and ($\Omega_c^*,\Sigma_c^*,\Xi^*_c$)
have been observed \cite{PDG}. Now we list out  all the charmed  baryon states from the particle data group.
The  $\Lambda^+_c$,
$\Lambda_c^+(2595)$, $\Lambda_c^+(2625)$, $\Lambda_c^+(2765)$ (or
$\Sigma_c^+(2765)$), $\Lambda_c^+(2880)$
  and $\Lambda_c^+(2940)$ have the spin-parity
   $J^P={\frac{1}{2}}^+$, ${\frac{1}{2}}^-$, ${\frac{3}{2}}^-$,  $?$, ${\frac{5}{2}}^+$ and $?$, respectively \cite{PDG}.
  The  $\Xi_c$, $\Xi^{\prime }_c$, $\Xi_c(2645)$,
$\Xi_c(2790)$, $\Xi_c(2815)$,
 $\Xi_c(2980)$,  $\Xi_c(3055)$, $\Xi_c(3080)$ and $\Xi_c(3123)$ have the spin-parity
   ${\frac{1}{2}}^+$,
  ${\frac{1}{2}}^+$,  ${\frac{3}{2}}^+$,  ${\frac{1}{2}}^-$,  ${\frac{3}{2}}^-$,
  $?$,  $?$,  $?$ and $?$,  respectively \cite{PDG}. The  $\Sigma_c(2455)$,
$\Sigma_c(2520)$ and $\Sigma_c(2800)$ have the spin-parity ${\frac{1}{2}}^+$,  ${\frac{3}{2}}^+$  and $?$,  respectively
\cite{PDG}. The $?$ denotes that the spin-parity is undetermined.

There have been several methods to study the heavy  baryon states, such as the QCD sum rules \cite{HeavyB-QCDSR,HeavyB-QCDSR-2008,HeavyB-QCDSR-2009,Wang0912,Wang0101,Wang0102,Wang1003,Wang0101-2,Wang1112}, the lattice QCD \cite{Latt-HB,Latt-HB-2014}, the relativistic
quark model \cite{HH-Ebert}, the relativized potential quark model \cite{Capstick1986},
the Feynman-Hellmann theorem \cite{Roncaglia95}, the combined expansion in $1/m_Q$ and $1/N_c$ \cite{Jenkins96},  the hyperfine
interaction \cite{Karliner09},  the variational approach \cite{HH-Roberts}, the Faddeev approach \cite{Valcarce08},  the unitarized  theory (or model) \cite{Garcia-Recio2013}, etc.

In Refs.\cite{ Wang0912,Wang0101,Wang0102,Wang1003,Wang0101-2,Wang1112},  we  study the  ${1\over 2}^{\pm}$ and ${3\over 2}^{\pm}$ heavy, doubly-heavy and triply-heavy
baryon states in a systematic way   with the QCD sum rules by subtracting the contributions from the corresponding ${1\over 2}^{\mp}$ and ${3\over 2}^{\mp}$  heavy, doubly-heavy and triply-heavy baryon states, and make reasonable predictions for their masses and pole residues.
For the heavy baryon states $\Lambda_c$ and $\Xi_c$, the predictions
$M_{\Lambda_c}=(2.26 \pm 0.27)\,\rm{GeV}$,
$M_{\Lambda_c(2595)}=(2.61 \pm 0.21)\,\rm{GeV}$,
 $M_{\Xi_c}=(2.44 \pm 0.23)\,\rm{GeV}$,
$M_{\Xi^{\prime}_c}=(2.56 \pm 0.22)\,\rm{GeV}$,
$M_{\Xi_c(2645)}=(2.65 \pm 0.20)\,\rm{GeV}$,
$M_{\Xi_c(2790)}=(2.76 \pm 0.18)\,\rm{GeV}$ and $M_{\Xi_c(2815)}=(2.86\pm0.17)\,\rm{GeV}$ are in good agreement
with the experimental data \cite{Wang0912,Wang0101,Wang0102,Wang1003}, where we take the $\Lambda_c(2595)$, $\Xi_c(2645)$, $\Xi_c(2790)$ and $\Xi_c(2815)\,\rm{GeV}$ to be  the $\Lambda$-type, $\Sigma$-type, $\Lambda$-type and $\Sigma$-type baryon states, respectively.
In the diquark-quark model for the baryons, if the two quarks in the diquark are in relative S-wave, then the baryons with the $0^+$ and $1^+$ diquarks (the ground state diquarks) are called $\Lambda$-type and $\Sigma$-type baryons respectively. On the other hand, if there exists  a relative P-wave between the two quarks in the diquark, then  the baryons with the $J^P=1^+\otimes 1^-$ and $0^+\otimes1^-$ diquarks are called $\Lambda$-type and $\Sigma$-type baryons respectively, where the $J^P=1^-$ denotes the relative P-wave, the $0^+$ and $1^+$ denote the spin-parity of the ground state diquarks.

The  flux-tube model favors to assign  the $\Lambda^+_c$,
$\Lambda_c^+(2595)$, $\Lambda_c^+(2625)$, $\Lambda_c^+(2765)$ (or
$\Sigma_c^+(2765)$), $\Lambda_c^+(2880)$
  and $\Lambda_c^+(2940)$ with the spin-parity  ${\frac{1}{2}}^+$, ${\frac{1}{2}}^-$,
  ${\frac{3}{2}}^-$,  ${\frac{3}{2}}^+$, ${\frac{5}{2}}^+$ and ${\frac{5}{2}}^-$, respectively \cite{Zhang2009}.
In the
non-relativistic quark model \cite{HH-Roberts}, the $\Xi_c(2790)$
  and $\Xi_c(2815)$ with the spin-parity ${\frac{1}{2}}^-$ and ${\frac{3}{2}}^-$
  respectively are assigned to be the charmed-strange analogues of
  the $\Lambda_c^+(2595)$ and $\Lambda_c^+(2625)$, or of the $\Lambda(1405)$ and
$\Lambda(1520)$; i.e. they are
  flavor antitriplet  or $\Lambda$-type heavy baryon states.
 In   the relativistic quark model \cite{HH-Ebert}, the $\Xi_c(2815)$ also is taken to be  the
$\Lambda$-type baryon state.

The  $\Xi_c(2815)$ may be the $\Lambda$-type or  $\Sigma$-type baryon state with the spin-parity
${\frac{3}{2}}^-$, there are two possibilities, while the  $\Xi_c(2980)$,
$\Xi_c(3055)$, $\Xi_c(3080)$ and $\Xi_c(3123)$ are unlikely the ground  state ${\frac{3}{2}}^-$ states due to their large masses.
In this article, we will focus on the possible assignments of the $\Lambda_c(2625)$ and $\Xi_c(2815)$ to be  the $\Lambda$-type baryon states.
In previous work, we take the $\Xi_c(2815)$ to be the $\Sigma$-type baryon state  \cite{Wang1003}.

We usually resort to the diquark-quark model
to construct the baryon currents. Without introducing additional P-wave, the ground state quarks have the spin-parity ${1\over 2}^+$, two
quarks can form a scalar diquark or an axialvector diquark with the
spin-parity $0^+$ or $1^+$, the
diquark then combines  with a third quark to form a positive parity
baryon,
\begin{eqnarray}
\left[{1\over 2}^+\otimes{1\over 2}^+\right]\otimes{1\over 2}^+&=&\left[{0}^{+} \oplus{1}^{+}\right] \otimes{1\over 2}^+={1\over 2}^{+}\oplus{1\over 2}^{+}\oplus{3\over 2}^+\, ,
\end{eqnarray}
for example, the $\Lambda$-type currents $\eta^\Lambda$,
\begin{eqnarray}
\eta^\Lambda&=&\varepsilon^{abc}q^T_aC\gamma_5 q^{\prime}_b \,Q_c,
\end{eqnarray}
the $\Sigma$-type currents $\eta^\Sigma$ and $\eta^\Sigma_\mu$,
\begin{eqnarray}
\eta^\Sigma&=&\varepsilon^{abc}q^T_aC\gamma_\mu q^{\prime}_b\,\gamma^\mu \gamma_5 Q_c \, , \nonumber\\
\eta^\Sigma_\mu&=&\varepsilon^{abc}q^T_aC\gamma_\mu q^{\prime}_b\,  Q_c \, ,
\end{eqnarray}
which have positive  parity, where the $a$, $b$ and $c$ are color indices. Multiplying $i \gamma_{5}$ to the currents $\eta^\Lambda$, $\eta^\Sigma$ and $\eta^\Sigma_\mu$ changes their parity, the currents $i \gamma_{5}\eta^\Lambda$, $i \gamma_{5}\eta^\Sigma$ and $i \gamma_{5}\eta^\Sigma_\mu$ couple potentially to the negative parity heavy baryons.
In Refs.\cite{Wang0101,Wang1003,Wang1112}, we take the currents without introducing partial (or P-wave) to study the negative parity heavy, doubly-heavy and triply-heavy   baryon states, and obtain satisfactory results.

If there exists a relative P-wave (which can be denoted as
$1^-$) between the diquark and the third quark or between the two quarks in the diquark, we have the following two routines to construct the negative parity
baryons,
\begin{eqnarray}
\left[{1\over 2}^+\otimes{1\over 2}^+\right]\otimes\left({1\over 2}^+\otimes{1^-}\right)&=&\left[{0}^{+} \oplus{1}^{+}\right] \otimes\left({1\over 2}^-\oplus{3\over 2}^-\right)
\, ,
\end{eqnarray}
and
\begin{eqnarray}
\left[\left({1\over 2}^+\otimes{1^-}\right)\otimes{1\over 2}^+\right]\otimes {1\over 2}^+&=&\left[\left({1\over 2}^-\oplus{3\over 2}^-\right)\otimes{1\over 2}^+\right]\otimes {1\over 2}^+  \nonumber\\
&=&\left[0^-\oplus 1^-\oplus 1^-\oplus2^- \right]\otimes {1\over 2}^+ \, ,
\end{eqnarray}
or equivalently
\begin{eqnarray}
\left[\left({1\over 2}^+\otimes{1\over 2}^+\right)\otimes{1^-}\right]\otimes {1\over 2}^+&=&\left[\left(0^+\oplus 1^+\right)\otimes 1^-\right]\otimes {1\over 2}^+  \nonumber\\
&=&\left[1^-\oplus 0^-\oplus 1^-\oplus2^- \right]\otimes {1\over 2}^+ \, .
\end{eqnarray}
Recently, Chen et al  introduce the relative P-wave explicitly, and study the negative parity charmed baryon states with the QCD sum rules combined with the heavy quark effective theory \cite{PwaveQCDSR}. The baryons have complicated structures, more than one currents can couple potentially to a special  baryon. In this article, we construct the interpolating currents by introducing the relative P-wave explicitly,   and study the  negative parity charmed baryon states $\Lambda_c(2625)$ and $\Xi_c(2815)$ with  the full QCD sum rules.

In Ref.\cite{Oka96}, Jido, Kodama and Oka suggest a novel  method to separate the contribution of   the
negative-parity  baryon $N(1535)$ from that of the positive-parity  baryon $p$, because the interpolating currents maybe
couple potentially to both the negative- and positive-parity baryon states
\cite{Chung82}, which impairs the  predictive power.
Again, we follow  this   novel  method  to
study the   negative-parity   baryon states $\Lambda_c(2625)$ and $\Xi_c(2815)$
by separating the contributions of  the positive-parity   baryon states explicitly. In the heavy quark limit,  Bagan et al  separate the
contributions of the positive- and negative-parity heavy baryon states
 unambiguously  \cite{Bagan93}.

 The article is arranged as follows:  we derive the
QCD sum rules for the masses and pole residues of  the
$\Lambda_c(2625)$ and $\Xi_c(2815)$
in Sect.2;
 in Sect.3, we present the numerical results and discussions; and Sect.4 is reserved for our
conclusions.

\section{QCD sum rules for  the $\Lambda_c(2625)$ and $\Xi_c(2815)$}

In the following, we write down  the two-point correlation functions $\Pi_{\alpha\beta}(p)$  in the QCD sum rules,
\begin{eqnarray}
\Pi_{\alpha\beta}(p)&=&i\int d^4x e^{ip \cdot x} \langle0|T\left\{J_{\alpha}(x)\bar{J}_{\beta}(0)\right\}|0\rangle \, ,
\end{eqnarray}
where $J_\alpha(x)=J^1_\alpha(x),\,J^2_\alpha(x)$,
\begin{eqnarray}
 J^1_{\alpha}(x)&=&i\varepsilon^{ijk} \left[ \partial^\mu q^T_i(x) C\gamma^\nu q^\prime_j(x)- q^T_i(x) C\gamma^\nu \partial^{\mu}q^\prime_j(x)\right]\left(\widetilde{g}_{\alpha\mu}\gamma_\nu-\widetilde{g}_{\alpha\nu}\gamma_\mu \right)\gamma_5 c_k(x) \, , \\
J^2_{\alpha}(x)&=&i\varepsilon^{ijk} \left[ \partial^\mu q^T_i(x) C\gamma^\nu q^\prime_j(x)- q^T_i(x) C\gamma^\nu \partial^{\mu}q^\prime_j(x)\right]\left(g_{\alpha\mu}\gamma_\nu+g_{\alpha\nu}\gamma_\mu-\frac{1}{2}g_{\mu\nu}\gamma_\alpha \right)\gamma_5 c_k(x) \, ,\nonumber\\
\end{eqnarray}
$\widetilde{g}_{\mu\nu}=g_{\mu\nu}-\frac{1}{4}\gamma_\mu\gamma_\nu$,
the $i$, $j$, $k$ are color indices, the $C$ is the charge conjugation matrix.
The light diquark constituents  $\varepsilon^{ijk} \left[ \partial^\mu q^T_i(x) C\gamma^\nu q^\prime_j(x)- q^T_i(x) C\gamma^\nu \partial^{\mu}q^\prime_j(x)\right]$
in the currents $J_\alpha$ have the same formula, i.e. they   have the two Lorentz indices $\mu$ and $\nu$, and couple  potentially to  the spin-1 or 2 diquarks.
 The Dirac matrixes $\widetilde{g}_{\alpha\mu}\gamma_\nu-\widetilde{g}_{\alpha\nu}\gamma_\mu $ and $g_{\alpha\mu}\gamma_\nu+g_{\alpha\nu}\gamma_\mu-\frac{1}{2}g_{\mu\nu}\gamma_\alpha$ are anti-symmetric and symmetric respectively when interchanging
  the indices $\mu$ and $\nu$, which are contracted with the corresponding indices in the diquark constituents,  so the diquark constituents in the currents $J_\alpha^1$ and $J^2_\alpha$ have the spins 1 and 2, respectively.
Furthermore, the   currents $J^1_{\alpha}$ and $J^2_{\alpha}$ both  have negative parity. We use the currents $J_\alpha$ with $q=u$ and $q^{\prime}=d$ ($q=u$ and $q^{\prime}=s$ or $q=d$ and $q^{\prime}=s$) to interpolate the $\Lambda_c(2625)$ ($\Xi_c(2815)$).

The currents $J_\alpha(0)$ couple potentially to the ${\frac{3}{2}}^-$ charmed  baryon states $B^{-}$,
\begin{eqnarray}
\langle 0| J_{\alpha} (0)|B^{-}(p)\rangle &=&\lambda_{-} U^{-}_\alpha(p,s) \, ,
\end{eqnarray}
the spinor $U^{-}_\alpha(p,s)$ satisfies the Rarita-Schwinger equation $(\not\!\!p-M_{-})U^{-}_\alpha(p)=0$ and the relations $\gamma^\alpha U^{-}_\alpha(p,s)=0$,
$p^\alpha U^{-}_\alpha(p,s)=0$. The currents also satisfy the relation $\gamma^\alpha J_\alpha(x)=0$, which is consistent with Eq.(10). On the other hand, the
currents also couple to the positive parity baryon states $B^{+}$,
 \begin{eqnarray}
\langle 0| J_{\alpha} (0)|B^{+}(p)\rangle &=&\lambda_{+}i\gamma_5 U^{+}_\alpha(p,s) \, ,
\end{eqnarray}
the spinors $U^{\pm}_\alpha(p,s)$ have analogous  properties and $\lambda_{+}\neq 0$.

 We  insert  a complete set  of intermediate baryon states with the
same quantum numbers as the current operators $J_\alpha(x)$ and
$i\gamma_5 J_\alpha(x)$ into the correlation functions
$\Pi_{\alpha\beta}(p)$ to obtain the hadronic representation
\cite{SVZ79,PRT85}. After isolating the pole terms of the lowest
states of the charmed  baryons, we obtain the
following results:
\begin{eqnarray}
   \Pi_{\alpha\beta}(p) & = & \lambda_{-}^2  {\!\not\!{p}+ M_{-} \over M_{-}^{2}-p^{2}  } \left(- g_{\alpha\beta}+\frac{\gamma_\alpha
\gamma_\beta}{3}+\frac{2p_\alpha p_\beta}{3M_{-}^2}-\frac{p_\alpha
\gamma_\beta-p_\beta \gamma_\alpha}{3M_{-}}
\right)+\nonumber\\
&&  \lambda_+^2 {\!\not\!{p} -M_{+} \over M^{2}_+ -p^{2} }\left(- g_{\alpha\beta}+\frac{\gamma_\alpha
\gamma_\beta}{3}+\frac{2p_\alpha p_\beta}{3M_{+}^2}-\frac{p_\alpha
\gamma_\beta-p_\beta \gamma_\alpha}{3M_{+}}
\right)  +\cdots \nonumber \\
&=&\Pi(p)\left(- g_{\alpha\beta}\right)+\cdots \, ,
    \end{eqnarray}
where the $M_{\pm}$ are the masses of the lowest states with the
 parity $\pm$ respectively, and the $\lambda_{\pm}$ are the
corresponding pole residues (or couplings). In this article, we
choose the tensor structure $g_{\mu\nu}$ for analysis. If we take $\vec{p} = 0$, then
\begin{eqnarray}
  \rm{limit}_{\epsilon\rightarrow0}\frac{{\rm Im}  \Pi(p_{0}+i\epsilon)}{\pi} & = &
    \lambda_{-}^{2} {\gamma_{0} + 1\over 2} \delta(p_{0} - M_{-})+
    \lambda_+^2 {\gamma_{0} - 1\over 2} \delta(p_{0} - M_+) +\cdots \nonumber \\
  & = & \gamma_{0} A(p_{0}) + B(p_{0})+\cdots \, ,
\end{eqnarray}
where
\begin{eqnarray}
  A(p_{0}) & = &{1 \over 2}\left[\lambda_-^{2} \delta(p_{0} -
  M_{-})+   \lambda_+^{2}
  \delta(p_{0} - M_+)  \right]  \, , \nonumber \\
   B(p_{0}) & = & {1 \over 2} \left[ \lambda_-^{2}
  \delta(p_{0} - M_-)  - \lambda_+^{2} \delta(p_{0} -
  M_{+})\right] \, ,
\end{eqnarray}
the  $A(p_{0}) + B(p_{0})$ and $A(p_{0}) - B(p_{0})$ contain the
contributions  from the negative-  and
positive-parity baryon states,  respectively \cite{Oka96}.

We  calculate the light quark parts of the correlation functions
 $\Pi_{\alpha\beta}(p)$ with the full light quark propagators   in the coordinate space and
use the momentum space expression for the $c$-quark propagator,
\begin{eqnarray}
S_{ij}(x)&=& \frac{i\delta_{ij}\!\not\!{x}}{ 2\pi^2x^4}
-\frac{\delta_{ij}m_q}{4\pi^2x^2}-\frac{\delta_{ij}\langle
\bar{q}q\rangle}{12} +\frac{i\delta_{ij}\!\not\!{x}m_q
\langle\bar{q}q\rangle}{48}-\frac{\delta_{ij}x^2\langle \bar{q}g_s\sigma Gq\rangle}{192}+\frac{i\delta_{ij}x^2\!\not\!{x} m_q\langle \bar{q}g_s\sigma
 Gq\rangle }{1152}\nonumber\\
&& -\frac{ig_s G^{a}_{\alpha\beta}t^a_{ij}(\!\not\!{x}
\sigma^{\alpha\beta}+\sigma^{\alpha\beta} \!\not\!{x})}{32\pi^2x^2} -\frac{1}{8}\langle\bar{q}_j\sigma^{\mu\nu}q_i \rangle \sigma_{\mu\nu} +\cdots \, ,
\end{eqnarray}
\begin{eqnarray}
C_{ij}(x)&=&\frac{i}{(2\pi)^4}\int d^4k e^{-ik \cdot x} \left\{
\frac{\delta_{ij}}{\!\not\!{k}-m_c}
-\frac{g_sG^n_{\alpha\beta}t^n_{ij}}{4}\frac{\sigma^{\alpha\beta}(\!\not\!{k}+m_c)+(\!\not\!{k}+m_c)
\sigma^{\alpha\beta}}{(k^2-m_c^2)^2}\right.\nonumber\\
&&\left.-\frac{g_s^2 (t^at^b)_{ij} G^a_{\alpha\beta}G^b_{\mu\nu}(f^{\alpha\beta\mu\nu}+f^{\alpha\mu\beta\nu}+f^{\alpha\mu\nu\beta}) }{4(k^2-m_c^2)^5}+\cdots\right\} \, ,\nonumber\\
f^{\alpha\beta\mu\nu}&=&(\!\not\!{k}+m_c)\gamma^\alpha(\!\not\!{k}+m_c)\gamma^\beta(\!\not\!{k}+m_c)\gamma^\mu(\!\not\!{k}+m_c)\gamma^\nu(\!\not\!{k}+m_c)\, ,
\end{eqnarray}
and  $t^n=\frac{\lambda^n}{2}$, the $\lambda^n$ is the Gell-Mann matrix  \cite{PRT85}. We contract the quark fields in the correlation functions and take the full light-quark and heavy-quark propagators firstly, then compute  the integrals both in the coordinate and momentum spaces,  and obtain the correlation functions $\Pi_{\alpha\beta}(p)$ therefore the QCD spectral densities  through  dispersion relation, the explicit expression are give in the appendix. In Eq.(15), we retain the term $\langle\bar{q}_j\sigma_{\mu\nu}q_i \rangle$  originates  from the Fierz re-arrangement of the $\langle q_i \bar{q}_j\rangle$ to  absorb the gluons  emitted from the other quark lines to form
$\langle\bar{q}_j g_s G^a_{\alpha\beta} t^a_{mn}\sigma_{\mu\nu} q_i \rangle$  so as to extract the mixed condensate  $\langle\bar{q}g_s\sigma G q\rangle$.
  Finally we introduce the weight
functions $\exp\left(-\frac{p_0^2}{T^2}\right)$,
$p_0^2\exp\left(-\frac{p_0^2}{T^2}\right)$,   and obtain the
following QCD sum rules,
\begin{eqnarray}
  \lambda_{-}^2\exp\left(-\frac{M_-^2}{T^2}\right)&=&\int_{m_c}^{\sqrt{s_0}}dp_0\left[\rho^A(p_0)+\rho^B(p_0)\right]\exp\left(-\frac{p_0^2}{T^2}\right) \, ,
\end{eqnarray}
\begin{eqnarray}
  \lambda_{-}^2M_-^2\exp\left(-\frac{M_-^2}{T^2}\right)&=&\int_{m_c}^{\sqrt{s_0}}dp_0\left[\rho^A(p_0)+\rho^B(p_0)\right]p_0^2\exp\left(-\frac{p_0^2}{T^2}\right) \, ,
\end{eqnarray}
where  the $s_0$ are the continuum threshold parameters and the $T^2$ are the
Borel parameters. The QCD spectral densities $\rho^A(p_0)$
and $\rho^B(p_0)$  are
given explicitly in the Appendix.

\section{Numerical results and discussions}
The vacuum condensates are taken to be the standard values
$\langle\bar{q}q \rangle=-(0.24\pm 0.01\, \rm{GeV})^3$,  $\langle\bar{s}s \rangle=(0.8\pm0.1)\langle\bar{q}q \rangle$,
$\langle\bar{q}g_s\sigma G q \rangle=m_0^2\langle \bar{q}q \rangle$, $\langle\bar{s}g_s\sigma G s \rangle=m_0^2\langle \bar{s}s \rangle$,
$m_0^2=(0.8 \pm 0.1)\,\rm{GeV}^2$, $\langle \frac{\alpha_s
GG}{\pi}\rangle=(0.33\,\rm{GeV})^4 $    at the energy scale  $\mu=1\, \rm{GeV}$
\cite{SVZ79,PRT85,Ioffe2005}.
The quark condensate and mixed quark condensate evolve with the   renormalization group equation,
$\langle\bar{q}q \rangle(\mu)=\langle\bar{q}q \rangle(Q)\left[\frac{\alpha_{s}(Q)}{\alpha_{s}(\mu)}\right]^{\frac{4}{9}}$,
 $\langle\bar{s}s \rangle(\mu)=\langle\bar{s}s \rangle(Q)\left[\frac{\alpha_{s}(Q)}{\alpha_{s}(\mu)}\right]^{\frac{4}{9}}$,
 $\langle\bar{q}g_s \sigma Gq \rangle(\mu)=\langle\bar{q}g_s \sigma Gq \rangle(Q)\left[\frac{\alpha_{s}(Q)}{\alpha_{s}(\mu)}\right]^{\frac{2}{27}}$
 and $\langle\bar{s}g_s \sigma Gs \rangle(\mu)=\langle\bar{s}g_s \sigma Gs \rangle(Q)\left[\frac{\alpha_{s}(Q)}{\alpha_{s}(\mu)}\right]^{\frac{2}{27}}$.

In the article, we take the $\overline{MS}$ masses $m_{c}(m_c)=(1.275\pm0.025)\,\rm{GeV}$ and $m_s(\mu=2\,\rm{GeV})=(0.095\pm0.005)\,\rm{GeV}$
 from the particle data group \cite{PDG}, and take into account
the energy-scale dependence of  the $\overline{MS}$ masses from the renormalization group equation,
\begin{eqnarray}
m_c(\mu)&=&m_c(m_c)\left[\frac{\alpha_{s}(\mu)}{\alpha_{s}(m_c)}\right]^{\frac{12}{25}} \, ,\nonumber\\
m_s(\mu)&=&m_s({\rm 2GeV} )\left[\frac{\alpha_{s}(\mu)}{\alpha_{s}({\rm 2GeV})}\right]^{\frac{4}{9}} \, ,\nonumber\\
\alpha_s(\mu)&=&\frac{1}{b_0t}\left[1-\frac{b_1}{b_0^2}\frac{\log t}{t} +\frac{b_1^2(\log^2{t}-\log{t}-1)+b_0b_2}{b_0^4t^2}\right]\, ,
\end{eqnarray}
  where $t=\log \frac{\mu^2}{\Lambda^2}$, $b_0=\frac{33-2n_f}{12\pi}$, $b_1=\frac{153-19n_f}{24\pi^2}$, $b_2=\frac{2857-\frac{5033}{9}n_f+\frac{325}{27}n_f^2}{128\pi^3}$,  $\Lambda=213\,\rm{MeV}$, $296\,\rm{MeV}$  and  $339\,\rm{MeV}$ for the flavors  $n_f=5$, $4$ and $3$, respectively  \cite{PDG}.

In Refs.\cite{WangTetraquark,WangTetraquark2015,WangMolecule}, we study the acceptable energy scales of the QCD spectral densities  for the hidden  charmed (bottom) tetraquark states and molecular (and molecule-like) states in the QCD sum rules in details for the first time,  and suggest a  formula $\mu=\sqrt{M^2_{X/Y/Z}-(2{\mathbb{M}}_Q)^2}$ to determine  the energy scales, where the $X$, $Y$, $Z$ denote the four-quark systems, and the ${\mathbb{M}}_Q$ is the effective heavy quark mass.
We can describe the  system $Q\bar{Q}q^{\prime}\bar{q}$ by a double-well potential with two light quarks $q^{\prime}\bar{q}$ lying in the two wells respectively.
   In the heavy quark limit, the $Q$-quark serves as a static well potential and bounds  the light quark $q^{\prime}$ to form a diquark in the color antitriplet channel or binds the light antiquark $\bar{q}$ to
form a meson (or meson-like) in the color singlet (or octet) channel.
Then the  four-quark systems  are characterized by the effective  masses ${\mathbb{M}}_Q$ and
the virtuality $V=\sqrt{M^2_{X/Y/Z}-(2{\mathbb{M}}_Q)^2}$.
We assume $ \mu^2=V^2={\mathcal{O}}(T^2)$,
 the effective mass ${\mathbb{M}}_c=1.8\,\rm{GeV}$ is the optimal value for  the diquark-antidiquark type tetraquark states \cite{WangTetraquark,WangTetraquark2015}. In this article, we use the diquark-quark model to construct the interpolating  currents, and take the analogous formula,
 \begin{eqnarray}
 \mu&=&\sqrt{M_{\Lambda_c/\Xi_c}^2-{\mathbb{M}}_c^2}\, ,
   \end{eqnarray}
   with  the value ${\mathbb{M}}_c=1.8\,\rm{GeV}$ to determine the energy scales of the QCD spectral densities. Then we obtain the values $\mu=1.9\,\rm{GeV}$ and $\mu=2.2\,\rm{GeV}$ for the $\Lambda_c(2625)$
 and $\Xi_c(2815)$, respectively.

In the conventional QCD sum rules \cite{SVZ79,PRT85}, we usually use  two
criteria (pole dominance and convergence of the operator product
expansion) to choose   the Borel parameters $T^2$ and continuum threshold
parameters $s_0$.  In Refs.\cite{ Wang0912,Wang0101,Wang0102,Wang1003,Wang0101-2,Wang1112},  we  study the  ${1\over 2}^{\pm}$ and ${3\over 2}^{\pm}$ heavy, doubly-heavy and triply-heavy baryon states in a systematic way  with the QCD sum rules by subtracting the contributions from the corresponding ${1\over 2}^{\mp}$ and ${3\over 2}^{\mp}$  heavy, doubly-heavy and triply-heavy baryon states, the continuum threshold parameters $\sqrt{s_0}-M_{\rm{gr}}\approx (0.6-0.8)\,\rm{GeV}$ can lead to satisfactory results, where $M_{\rm{gr}}$ denotes the ground state masses.
The masses of the $\Lambda_c(2625)$ and $\Xi_c(2815)$ are $M_{\Lambda_c(2625)}=(2628.11\pm0.19)\,\rm{MeV}$, $M_{\Xi^+_c(2815)}=(2816.6\pm0.9)\,\rm{MeV}$ and $M_{\Xi^0_c(2815)}=(2819.6\pm1.2)\,\rm{MeV}$ from the particle data group \cite{PDG}.
In this article,  we take the values
$\sqrt{s_0}\approx M_{\rm gr}+(0.6-0.8)\,\rm{GeV}$,   the two criteria of the QCD sum rules are also satisfied, see Table 1.
 The values $(0.6-0.8)\,\rm{GeV}$ are somewhat larger than the usually used values $(0.4-0.6)\,\rm{GeV}$, there maybe exist some contaminations from the higher resonances. If we take the largest values $\sqrt{s_0}= M_{\rm gr}+0.8\,\rm{GeV}$, the upper bound of the factors $\exp\left(-\frac{s_0}{T^2} \right)$ is about $0.003-0.005$, the contaminations are greatly suppressed and can be neglected safely.
  In the table, we present the values of the Borel parameters $T^2$, continuum threshold parameters $s_0$, the pole contributions and the perturbative contributions explicitly.

\begin{table}
\begin{center}
\begin{tabular}{|c|c|c|c|c|c|}\hline\hline
                                      & $T^2 (\rm{GeV}^2)$   & $\sqrt{s_0} (\rm{GeV})$    & pole          & perturbative \\ \hline
$\Lambda_c(2625)\,\,(J^1_\alpha)$     & $1.6-2.0$            & $3.3$                      & $(50-72)\%$   & $(81-95)\%$ \\ \hline
$\Lambda_c(2625)\,\,(J^2_\alpha)$     & $1.8-2.2$            & $3.3$                      & $(45-65)\%$   & $(76-88)\%$ \\ \hline
$\Xi_c(2815)\,\,(J^1_\alpha)$         & $1.6-2.2$            & $3.5$                      & $(54-82)\%$   & $\geq 89\%$ \\ \hline
$\Xi_c(2815)\,\,(J^2_\alpha)$         & $1.8-2.4$            & $3.5$                      & $(50-75)\%$   & $(82-94)\%$ \\ \hline
\end{tabular}
\end{center}
\caption{ The Borel parameters $T^2$, continuum threshold parameters $s_0$,
the pole contributions (pole)   and the perturbative contributions (perturbative).}
\end{table}

Taking into account all uncertainties  of the revelent  parameters,
we can obtain the values of the masses and pole residues of
 the  $\Lambda_c(2625)$ and $\Xi_c(2815)$, which are shown in Figs.1-2 and
Table 2. From the table, we can see that the values of the masses $M_{\Lambda_c(2625)}$ and $M_{\Xi_c(2815)}$ can reproduce the experimental data for all the currents $J_\alpha^1$ and $J_\alpha^2$.  The angular momentums  of the light diquarks are $1$ and $2$ in the currents $J_\alpha^1$ and $J_\alpha^2$, respectively,  they all couple potentially to the baryons  $\Lambda_c(2625)$ and $\Xi_c(2815)$, so the $\Lambda_c(2625)$ and $\Xi_c(2815)$ have at least two remarkable under-structures.

In previous work \cite{Wang1003}, we take the $\Xi_c(2815)$ to be the $\Sigma$-type baryon state,
and study the $\Xi_c(2815)$ with the interpolating current  $J_\alpha^{\Xi}(x)= \epsilon^{ijk}  q^T_i(x)C\gamma_\alpha s_j(x)    c_k(x)$ or $J_\alpha^{\Xi}(x)=\epsilon^{ijk}  q^T_i(x)C\gamma^\beta s_j(x) \widetilde{g}_{\alpha\beta}   c_k(x)$,
and obtain the value
$M_{\Xi_c(2815)}=(2.86\pm0.17)\,\rm{GeV}$, which is also  consistent
with the experimental data. If the prediction is robust, now the $\Xi_c(2815)$ has at least three remarkable under-structures.

In Fig.3, we plot the masses $M_{\Lambda_c(2625)}$ and $M_{\Xi_c(2815)}$ with variations of the energy scales  $\mu$ for the central values of the other input parameters.
From the figure, we can see that the $M_{\Lambda_c(2625)}$  and $M_{\Xi_c(2815)}$ decrease  monotonously but mildly with increase of the energy scales $\mu$, $M_{\Lambda_c(2625)}\approx (2.60-2.63)\,\rm{GeV}$ and  $M_{\Xi_c(2815)}\approx (2.82-2.88)\,\rm{GeV}$ at the energy scales $\mu=(1-3)\,\rm{GeV}$, the allowed energy scales are $\mu_{\Lambda_c(2625)}=(1-3)\,\rm{GeV}$ and $\mu_{\Xi_c(2815)}=(1.4-3.0)\,\rm{GeV}$, if we assume $M_{\Xi_c(2815)}\leq 2.86\,\rm{GeV}$, so the energy scale formula in Eq.(20) works, the formula can be extend to study other heavy baryon states.

\begin{table}
\begin{center}
\begin{tabular}{|c|c|c|c|c|c|c|}\hline\hline
                                    & $T^2 (\rm{GeV}^2)$  & $\sqrt{s_0} (\rm{GeV})$   & $M(\rm{GeV})$   &$\lambda (\rm{GeV}^4)$    \\ \hline
 $\Lambda_c(2625)\,\,(J^1_\alpha)$  & $1.6-2.0$           & $3.3\pm0.1$               & $2.62\pm0.18$   &$0.041\pm0.014$            \\ \hline
 $\Lambda_c(2625)\,\,(J^2_\alpha)$  & $1.8-2.2$           & $3.3\pm0.1$               & $2.61\pm0.18$   &$0.072\pm0.022$            \\ \hline
 $\Xi_c(2815)\,\,(J^1_\alpha)$      & $1.6-2.2$           & $3.5\pm0.1$               & $2.83\pm0.17$   &$0.065\pm0.022$            \\ \hline
 $\Xi_c(2815)\,\,(J^2_\alpha)$      & $1.8-2.4$           & $3.5\pm0.1$               & $2.83\pm0.17$   &$0.113\pm0.034$            \\ \hline
    \hline
\end{tabular}
\end{center}
\caption{ The masses $M$ and pole residues
$\lambda$ of the $\Lambda_c(2625)$ and $\Xi_c(2815)$.}
\end{table}

\begin{figure}
 \centering
 \includegraphics[totalheight=5cm,width=6cm]{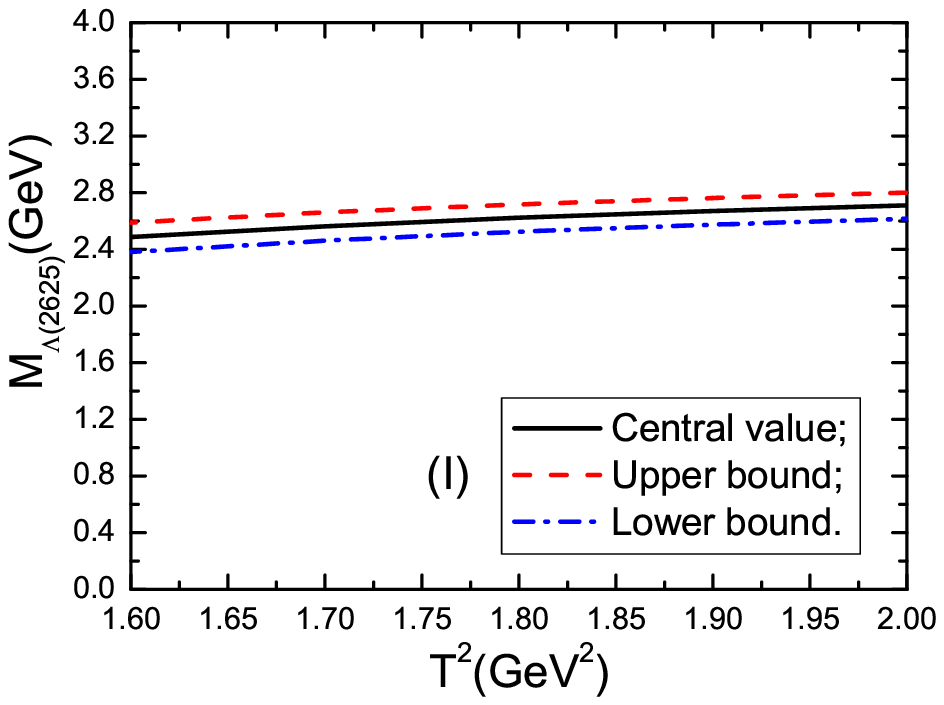}
 \includegraphics[totalheight=5cm,width=6cm]{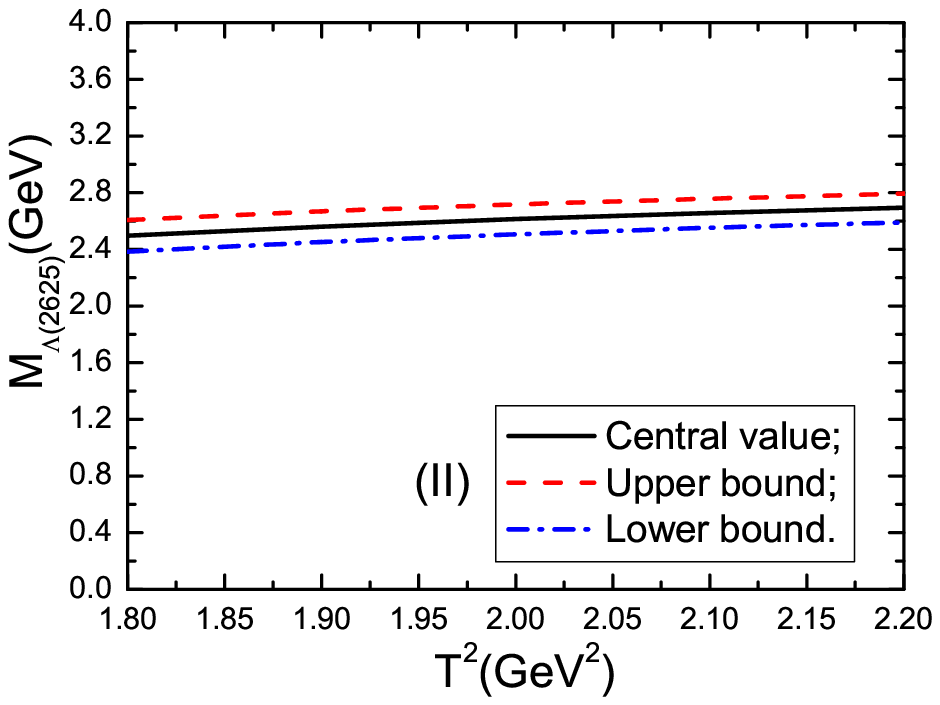}
 \includegraphics[totalheight=5cm,width=6cm]{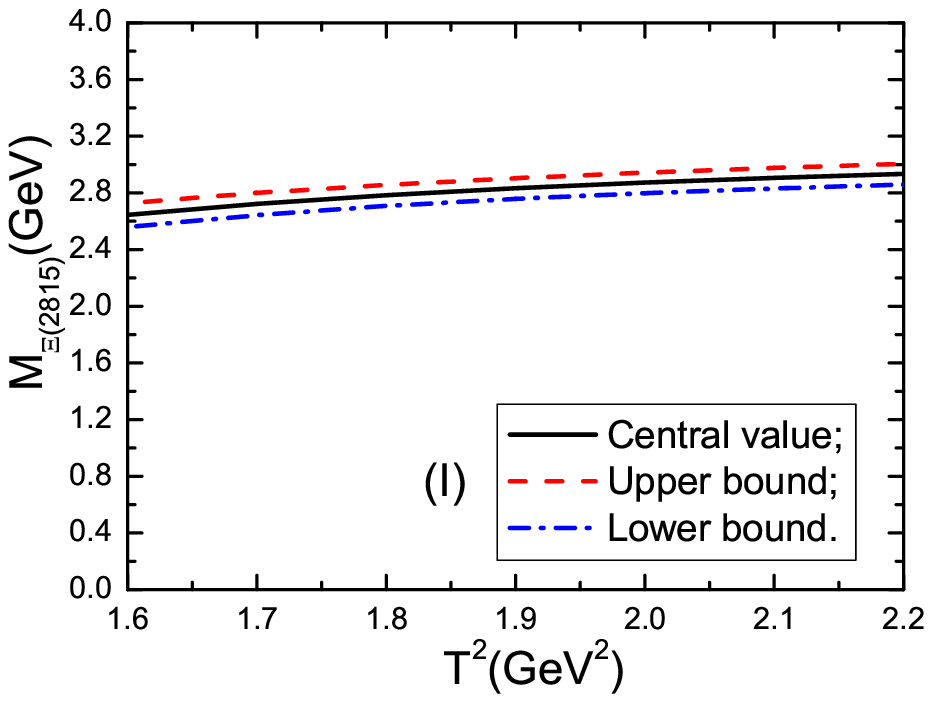}
 \includegraphics[totalheight=5cm,width=6cm]{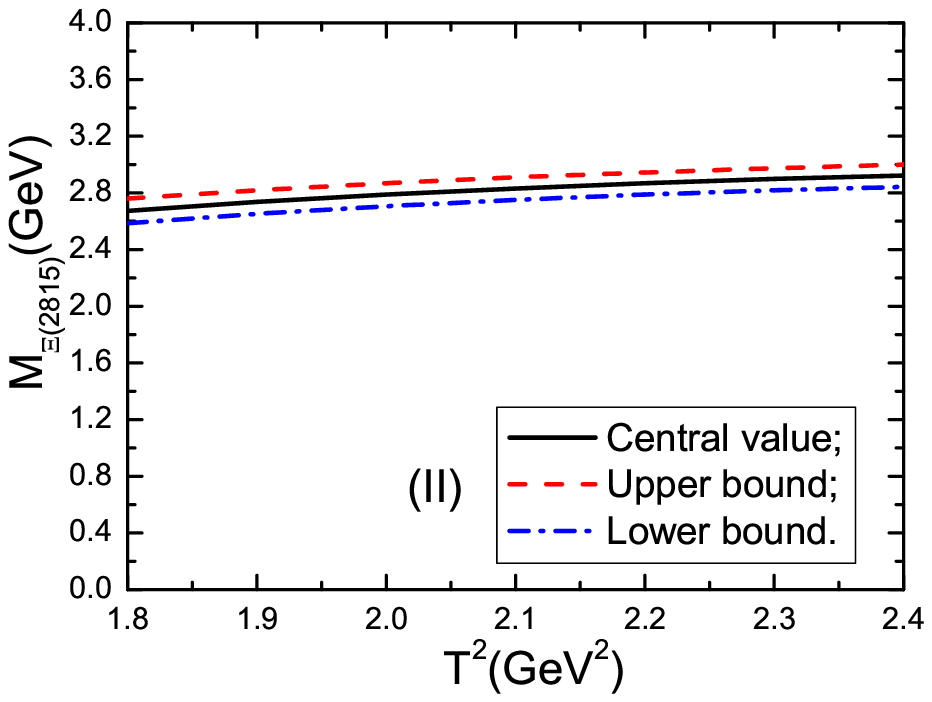}
       \caption{ The masses  of the  $\Lambda_c(2625)$ and $\Xi_c(2815)$  with variations of the Borel parameters $T^2$, where the (I) and (II) denote the currents
        $J^1_\alpha$ and $J^2_\alpha$, respectively.  }
\end{figure}

\begin{figure}
 \centering
 \includegraphics[totalheight=5cm,width=6cm]{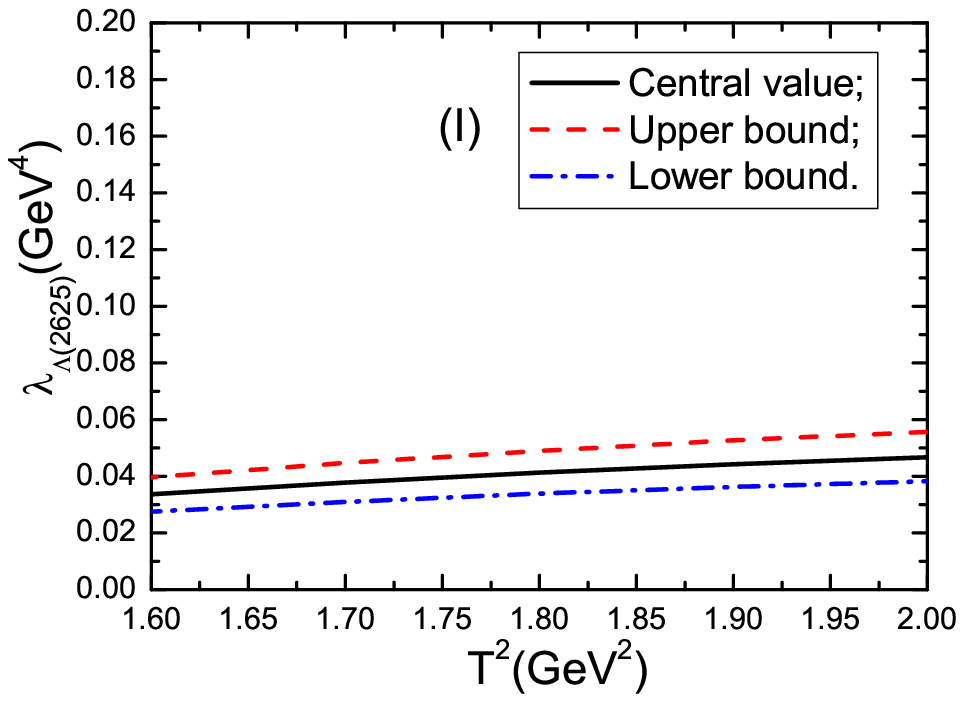}
 \includegraphics[totalheight=5cm,width=6cm]{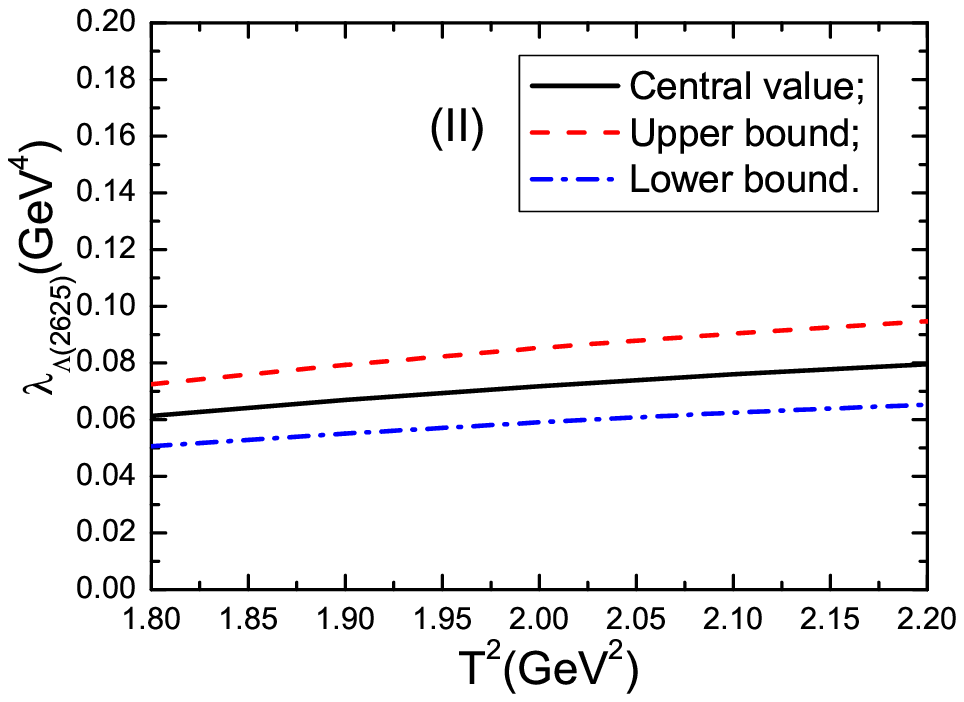}
 \includegraphics[totalheight=5cm,width=6cm]{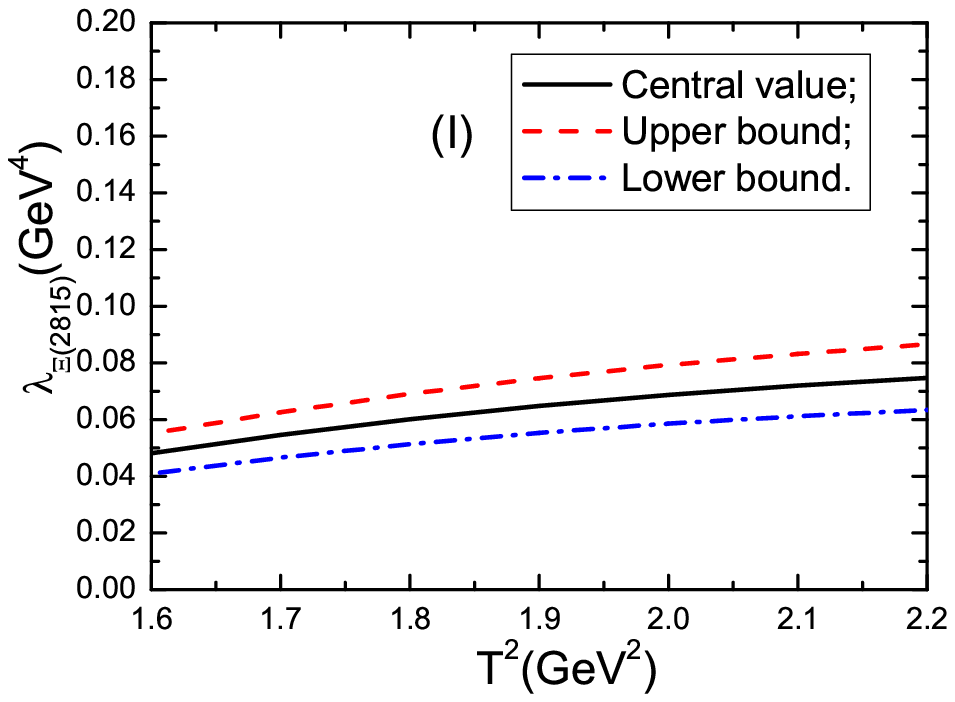}
 \includegraphics[totalheight=5cm,width=6cm]{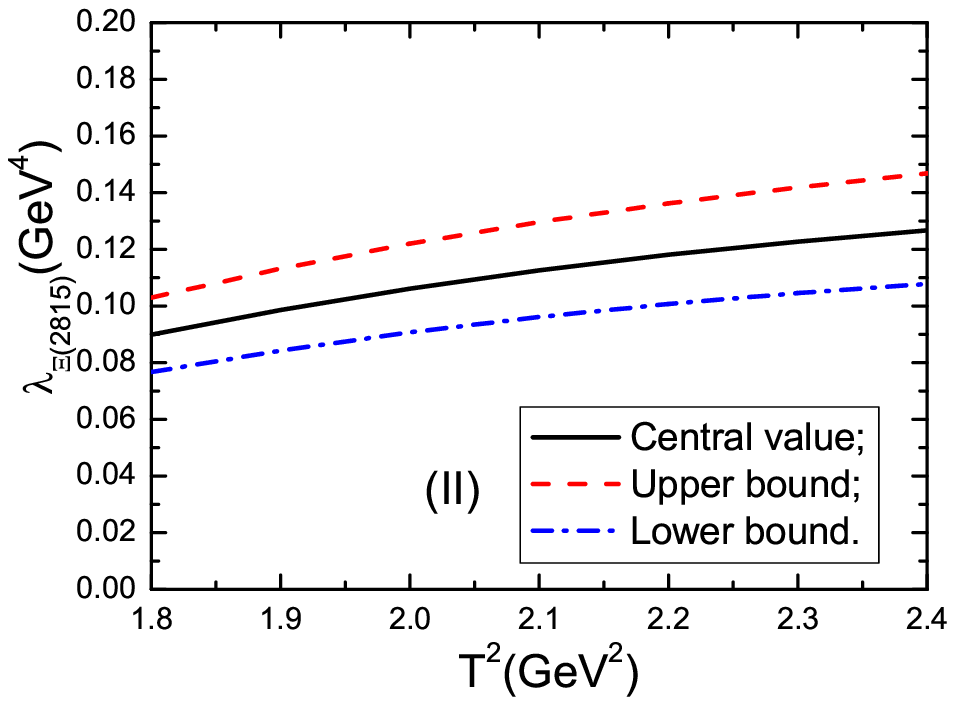}
       \caption{ The pole residues   of the  $\Lambda_c(2625)$ and $\Xi_c(2815)$  with variations of the Borel parameter $T^2$, where the (I) and (II) denote the currents
        $J^1_\alpha$ and $J^2_\alpha$, respectively.  }
\end{figure}

\begin{figure}
 \centering
 \includegraphics[totalheight=5cm,width=6cm]{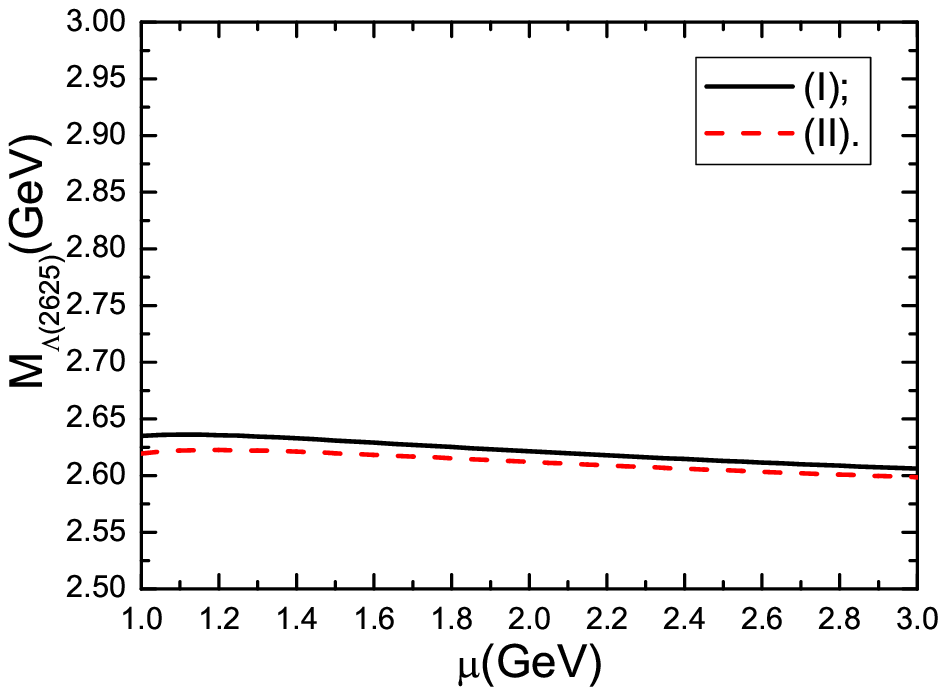}
 \includegraphics[totalheight=5cm,width=6cm]{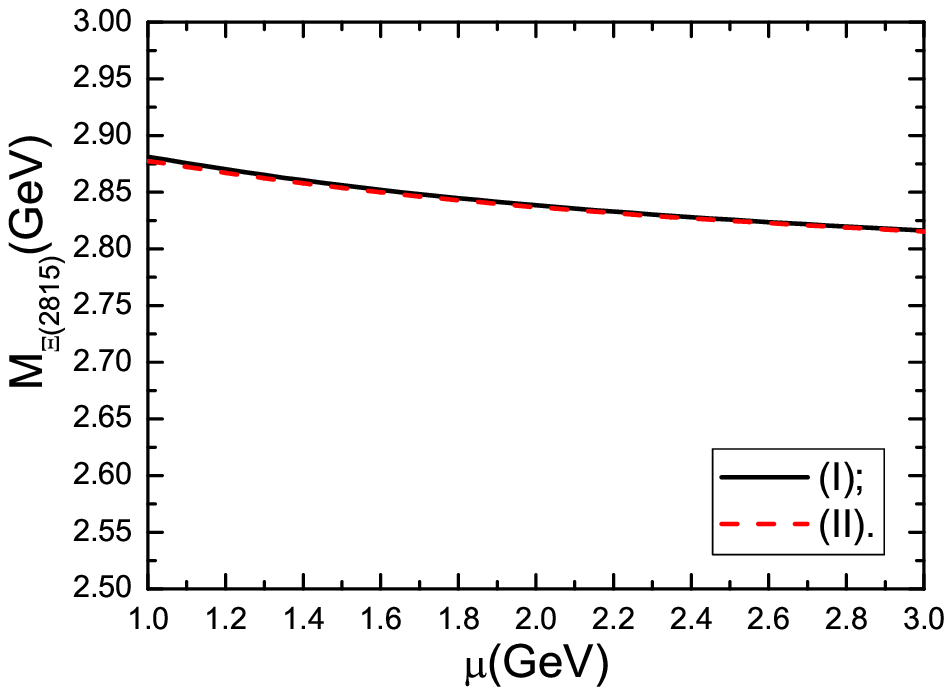}
       \caption{ The masses  of the  $\Lambda_c(2625)$ and $\Xi_c(2815)$  with variations of the energy scales  $\mu$ where the (I) and (II) denote the currents
        $J^1_\alpha$ and $J^2_\alpha$, respectively.  }
\end{figure}

\section{Conclusion}
In this article, we study   the
charmed baryon states $\Lambda_c(2625)$ and $\Xi_c(2815)$ with the spin-parity ${\frac{3}{2}^-}$
 by subtracting the contributions
from the corresponding    charmed  baryon states with the spin-parity ${\frac{3}{2}^+}$ using the QCD sum rules, and suggest an energy scale formula to determine the energy scales of the QCD spectral densities, and make reasonable predictions for their masses and pole residues.  The numerical
results  indicate that the  $\Lambda_c(2625)$ and $\Xi_c(2815)$ at least have two remarkable under-structures. We can take pole residues  as   basic input parameters and study the revelent hadronic processes with the QCD sum rules in further investigations of the under-structures of the $\Lambda_c(2625)$ and $\Xi_c(2815)$.

\section*{Acknowledgements}
This  work is supported by National Natural Science Foundation,
Grant Numbers 11375063, and Natural Science Foundation of Hebei province, Grant Number A2014502017.

\section*{Appendix}
The  spectral densities of the $\Lambda_c(2625)$ and $\Xi_c(2815)$
at the quark level,

\begin{eqnarray}
\rho^A_{J^1_{us}}(p_0)&=&\frac{p_0}{192\pi^4}\int_{t_i}^1dt
(1-t)^4\left(p_0^2-\widetilde{m}_c^2\right)^2\left[(4-5t)p_0^2+(2t-1)\widetilde{m}_c^2\right] \nonumber\\
&&+\frac{m_s\left[\langle\bar{q}q\rangle-2\langle\bar{s}s\rangle\right]p_0}{16\pi^2}\int_{t_i}^1 dt
(1-t)^2\left[(6t-5)p_0^2+(3-4t)\widetilde{m}_c^2\right]+\nonumber\\
&&\frac{m_s\left[12\langle\bar{q}g_s\sigma Gq\rangle-11\langle\bar{s}g_s\sigma Gs\rangle\right]p_0}{384\pi^2}\int_{t_i}^1dt
(1-t)\left[(7-8t)+2(1-t)p_0\delta(p_0-\widetilde{m}_c)\right] \nonumber\\
&& +\frac{p_0m_c^2}{576\pi^2}\langle \frac{\alpha_sGG}{\pi}\rangle
\int_{t_i}^1 dt \frac{(1-t)^4}{t^3}\left[2t-1-\frac{(1-t)p_0}{2}\delta(p_0-\widetilde{m}_c)\right]\nonumber \\
&& +\frac{m_s\langle\bar{q}g_s\sigma Gq\rangle p_0}{192\pi^2}\int_{t_i}^1 dt \frac{(1-t)^2}{t}\left[3-4t+\frac{(1-t)p_0}{2}\delta(p_0-\widetilde{m}_c)\right]\nonumber \\
&& +\frac{p_0}{768\pi^2}\langle \frac{\alpha_sGG}{\pi}\rangle
\int_{t_i}^1 dt \frac{(1-t)^3}{t}\left[(10t-7)p_0^2+(5-8t)\widetilde{m}_c^2\right]\, ,
\end{eqnarray}

\begin{eqnarray}
\rho^B_{J^1_{us}}(p_0)&=&\frac{m_c}{128\pi^4}\int_{t_i}^1dt
(1-t)^3\left(p_0^2-\widetilde{m}_c^2\right)^3+\frac{m_sm_c\left[5\langle\bar{s}g_s\sigma Gs\rangle-12\langle\bar{q}g_s\sigma Gq\rangle\right]}{128\pi^2}\int_{t_i}^1
dt \nonumber\\
&&+\frac{ \langle\bar{q} q\rangle\langle\bar{s}g_s\sigma Gs\rangle+\langle\bar{s}s\rangle\langle\bar{q}g_s\sigma Gq\rangle }{32}\delta(p_0-m_c) \nonumber\\
&&+\frac{m_c}{384\pi^2}\langle \frac{\alpha_sGG}{\pi}\rangle
\int_{t_i}^1 dt \frac{(1-t)^3}{t^2}\left( 3p_0^2-4\widetilde{m}_c^2\right)\nonumber \\
&& +\frac{m_sm_c\langle\bar{q}g_s\sigma Gq\rangle }{192\pi^2}\int_{t_i}^1 dt (t-1)\nonumber \\
&& -\frac{m_c}{2304\pi^2}\langle \frac{\alpha_sGG}{\pi}\rangle
\int_{t_i}^1 dt \frac{(1-t)^2(2t+1)}{t}\left(p_0^2-\widetilde{m}_c^2\right)\nonumber \\
&& -\frac{m_c}{128\pi^2}\langle \frac{\alpha_sGG}{\pi}\rangle
\int_{t_i}^1 dt (1-t)\left(p_0^2-\widetilde{m}_c^2\right) \, ,
\end{eqnarray}

\begin{eqnarray}
\rho^A_{J^2_{us}}(p_0)&=&\frac{p_0}{960\pi^4}\int_{t_i}^1dt
(1-t)^4\left(p_0^2-\widetilde{m}_c^2\right)^2\left[(42-9t-28t^2)p_0^2+(16t^2+3t-24)\widetilde{m}_c^2\right] \nonumber\\
&&+\frac{m_s\langle\bar{s}s\rangle p_0}{8\pi^2}\int_{t_i}^1 dt
(1-t)^2\left[(20t-20t^2-1)p_0^2+(16t^2-16t+1)\widetilde{m}_c^2\right]\nonumber\\
&&+\frac{m_s\langle\bar{q}q\rangle p_0}{8\pi^2}\int_{t_i}^1 dt
(1-t)^3\left(5\widetilde{m}_c^2-7p_0^2\right)\nonumber\\
&&+\frac{5m_s\langle\bar{s}g_s\sigma Gs\rangle p_0}{384\pi^2}\int_{t_i}^1dt
(1-t)\left(128t^2-180t+57 \right) \nonumber\\
&&+\frac{m_s\langle\bar{s}g_s\sigma Gs\rangle p_0^2}{96\pi^2}\int_{t_i}^1dt
(1-t)^2\left(9-20t \right)\delta(p_0-\widetilde{m}_c) \nonumber\\
&&+\frac{5m_s\langle\bar{q}g_s\sigma Gq\rangle p_0}{32\pi^2}\int_{t_i}^1dt
(1-t)\left(3-4t \right) \nonumber\\
&&+\frac{m_s\langle\bar{q}g_s\sigma Gq\rangle p_0^2}{8\pi^2}\int_{t_i}^1dt
(1-t)^2 \delta(p_0-\widetilde{m}_c) \nonumber\\
&&+\frac{5\left[ \langle\bar{q} q\rangle\langle\bar{s}g_s\sigma Gs\rangle+\langle\bar{s}s\rangle\langle\bar{q}g_s\sigma Gq\rangle\right] }{96}\delta(p_0-m_c) \nonumber\\
&& +\frac{p_0m_c^2}{2880\pi^2}\langle \frac{\alpha_sGG}{\pi}\rangle
\int_{t_i}^1 dt \frac{(1-t)^4}{t^3}\left(16t^2+3t-24 \right)\nonumber \\
&& -\frac{p_0^2m_c^2}{2880\pi^2}\langle \frac{\alpha_sGG}{\pi}\rangle
\int_{t_i}^1 dt \frac{(1-t)^5(2t+3)}{t^3} \delta(p_0-\widetilde{m}_c) \nonumber \\
&& -\frac{m_s\langle\bar{q}g_s\sigma Gq\rangle p_0}{192\pi^2}\int_{t_i}^1 dt \frac{(1-t)^2}{t}\left[1+2t+\frac{(1-t)p_0}{2}\delta(p_0-\widetilde{m}_c)\right]\nonumber \\
&& +\frac{p_0}{1152\pi^2}\langle \frac{\alpha_sGG}{\pi}\rangle
\int_{t_i}^1 dt \frac{(1-t)^3}{t}\left[(23+21t-20t^2)p_0^2+4(4t^2-5t-5)\widetilde{m}_c^2\right] \nonumber \\
&& +\frac{p_0}{1152\pi^2}\langle \frac{\alpha_sGG}{\pi}\rangle
\int_{t_i}^1 dt (1-t)^2\left[(68t-55-40t^2)p_0^2+(32t^2-52t+47)\widetilde{m}_c^2\right] \, , \nonumber \\
\end{eqnarray}

\begin{eqnarray}
\rho^B_{J^2_{us}}(p_0)&=&\frac{m_c}{192\pi^4}\int_{t_i}^1dt
(1-t)^3(t+4)\left(p_0^2-\widetilde{m}_c^2\right)^3\nonumber\\
&&+\frac{m_sm_c\left[\langle\bar{s}s\rangle-2\langle\bar{q}q\rangle\right]}{8\pi^2}\int_{t_i}^1 dt
t(1-t)\left(p_0^2-\widetilde{m}_c^2\right)\nonumber\\
&&+\frac{m_sm_c\left[17\langle\bar{s}g_s\sigma Gs\rangle-60\langle\bar{q}g_s\sigma Gq\rangle\right]}{384\pi^2}\int_{t_i}^1dt \nonumber\\
&&+\frac{m_sm_c\left[\langle\bar{s}g_s\sigma Gs\rangle-12\langle\bar{q}g_s\sigma Gq\rangle\right]}{48\pi^2}\int_{t_i}^1dt(1-t) \nonumber\\
&&+\frac{3\left[ \langle\bar{q} q\rangle\langle\bar{s}g_s\sigma Gs\rangle+\langle\bar{s}s\rangle\langle\bar{q}g_s\sigma Gq\rangle\right] }{32}\delta(p_0-m_c) \nonumber\\
&&+\frac{m_c}{576\pi^2}\langle \frac{\alpha_sGG}{\pi}\rangle
\int_{t_i}^1 dt \frac{(1-t)^3(t+4)}{t^2}\left( 3p_0^2-4\widetilde{m}_c^2\right)\nonumber \\
&& +\frac{7m_sm_c\langle\bar{q}g_s\sigma Gq\rangle }{192\pi^2}\int_{t_i}^1 dt (t-1)\nonumber \\
&& +\frac{m_c}{384\pi^2}\langle \frac{\alpha_sGG}{\pi}\rangle
\int_{t_i}^1 dt \frac{(1-t)^2(7t+11)}{t}\left(p_0^2-\widetilde{m}_c^2\right)\nonumber \\
&& -\frac{m_c}{384\pi^2}\langle \frac{\alpha_sGG}{\pi}\rangle
\int_{t_i}^1 dt (1-t)(2t+15)\left(p_0^2-\widetilde{m}_c^2\right) \, ,
\end{eqnarray}

\begin{eqnarray}
\rho^A_{J^1_{ud}}(p_0)&=& \rho^A_{J^1_{us}}(p_0)\mid_{m_s\to 0, \,\, \langle\bar{s}s\rangle \to \langle\bar{q}q\rangle, \,\,\langle\bar{s}g_s\sigma Gs\rangle \to \langle\bar{q}g_s\sigma Gq\rangle } \, , \nonumber\\
\rho^A_{J^2_{ud}}(p_0)&=& \rho^A_{J^2_{us}}(p_0)\mid_{m_s\to 0, \,\, \langle\bar{s}s\rangle \to \langle\bar{q}q\rangle, \,\,\langle\bar{s}g_s\sigma Gs\rangle \to \langle\bar{q}g_s\sigma Gq\rangle } \, , \nonumber\\
\rho^B_{J^1_{ud}}(p_0)&=& \rho^B_{J^1_{us}}(p_0)\mid_{m_s\to 0, \,\, \langle\bar{s}s\rangle \to \langle\bar{q}q\rangle, \,\,\langle\bar{s}g_s\sigma Gs\rangle \to \langle\bar{q}g_s\sigma Gq\rangle } \, , \nonumber\\
\rho^B_{J^2_{ud}}(p_0)&=& \rho^B_{J^2_{us}}(p_0)\mid_{m_s\to 0, \,\, \langle\bar{s}s\rangle \to \langle\bar{q}q\rangle, \,\,\langle\bar{s}g_s\sigma Gs\rangle \to \langle\bar{q}g_s\sigma Gq\rangle } \, ,
\end{eqnarray}
$\widetilde{m}_c^2=\frac{m_c^2}{t}$,
$t_i=\frac{m_c^2}{p_0^2}$, and we add the indices $us$ and $ud$ to denote the light quark constituents.

\end{document}